\documentclass[%
aip,
rsi,%
amsmath,amssymb,
reprint,%
]{revtex4-1}

\usepackage{graphicx}
\usepackage{dcolumn}
\usepackage{bm}
\usepackage{algorithm}%
\usepackage{algorithmicx}%
\usepackage{algpseudocode}%
\usepackage{float}
\usepackage[colorlinks=true, linkcolor=blue, citecolor=blue, urlcolor=blue]{hyperref}

\begin{document}
	
	\title{Principles and Metrics of Extreme Learning Machines Using a Highly Nonlinear Fiber}
	
\author{Mathilde Hary}
\address{Photonics Laboratory, Tampere University, FI-33104 Tampere, Finland}
\address{Universit\'{e} Marie et Louis Pasteur, CNRS UMR 6174, institut FEMTO-ST, 25000, Besan\c{c}on, France}
\email{mathilde.hary@tuni.fi}

\author{Daniel Brunner}
\address{Universit\'{e} Marie et Louis Pasteur, CNRS UMR 6174, institut FEMTO-ST, 25000, Besan\c{c}on, France}

\author{Lev Leybov}
\address{Photonics Laboratory, Tampere University, FI-33104 Tampere, Finland}

\author{Piotr Ryczkowski}
\address{Photonics Laboratory, Tampere University, FI-33104 Tampere, Finland}

\author{John M. Dudley}
\address{FEMTO-ST Institute/Optics Department, CNRS - University Franche-Comté, 15B avenue des Montboucons, Besançon Cedex, 25030, France}

\author{Go\"{e}ry Genty}
\address{Photonics Laboratory, Tampere University, FI-33104 Tampere, Finland}	
	
	\date{\today}
	
	\begin{abstract}
		
Optical computing offers potential for ultra high-speed and low-latency computation by leveraging the intrinsic properties of light, such as parallelism and linear as well as nonlinear ultra-high bandwidth signal transformations.
 Here, we explore the use of highly nonlinear optical fibers (HNLFs) as platforms for optical computing based on the concept of Extreme Learning Machines (ELMs).
 To evaluate the information processing potential of the system, we consider both task-independent and task-dependent performance metrics. The former focuses on intrinsic properties such as effective dimensionality, quantified via principal component analysis (PCA) on the system response to random inputs.
 The latter evaluates classification task accuracy on the MNIST digit dataset, highlighting how the system performs under different compression levels and nonlinear propagation regimes.
 We show that input power and fiber characteristics significantly influence the dimensionality of the computational system, with longer fibers and higher dispersion producing up to 100 principal components (PCs) at input power levels of 30 mW, where the PC corresponds to the linearly independent dimensions of the system.
 The spectral distribution of the PC's eigenvectors reveals that the high-dimensional dynamics facilitating computing through dimensionality expansion are located within 40~nm of the pump wavelength at 1560~nm, providing general insight for computing with nonlinear Schr\"odinger equation systems.  
 Task-dependent results demonstrate the effectiveness of HNLFs in classifying MNIST dataset images.
 Using input data compression through PC analysis, we inject MNIST images of various input dimensionality into the system and study the impact of input power upon classification accuracy.
 At optimized power levels, we achieve a classification test accuracy of 87\% ± 1.3 \%, significantly surpassing the baseline of 83.7\% from linear systems.
 Noteworthy, we find that the best performance is not obtained at maximal input power, i.e., maximal system dimensionality, but at more than one order of magnitude lower.
 The same is confirmed regarding the MNIST image's compression, where accuracy is substantially improved when strongly compressing the image to less than 50 PCs.
 These are highly relevant findings for the dimensioning of future, ultrafast optical computing systems that can capture and process sequential input information on femtosecond timescales.

	\end{abstract}
	
	\maketitle
	
\section{Introduction} 
\vspace*{-3pt}
Optical computing has emerged as a promising model for addressing the growing demands of high-speed and energy-efficient computation \cite{Shastri2021}.
 By leveraging the unique properties of light, such as high bandwidth and parallelism, optical systems offer significant advantages over traditional electronic architectures.
 This is particularly true for tasks involving real-time computing of ultrafast phenomena and ultrafast metrology \cite{Hall2001}.
 For such applications, electronics impose a GHz bandwidth limitation, which is elegantly mitigated using photonics.
 Computing approaches inspired by neural networks require a large number of linear connections and nonlinear transformations that preserve the time scale of the input data sample. 
 The nonlinear dynamics of photonic systems, and in particular those associated with the propagation of light in optical fiber, have the potential to significantly expand the horizon of real-time computing into the THz bandwidth domain.
 Only a limited number of optical systems have demonstrated practical and trainable computing behavior, with most implementations remaining proof-of-concept demonstrations and requiring various digital pre and post-processing steps. Current approaches are often constrained to specific architectures, and operate in the MHz to GHz range, far from the intrinsic THz bandwidths enabled by femtosecond lasers. Therefore, they do not fully exploit the ultrafast temporal resolution and nonlinear transformation potential offered by broadband nonlinear optical systems.

Artificial Neural Networks (ANNs) have become a cornerstone of modern computation due to their ability to process high-dimensional data effectively.
 The inherent parallelism of light has the potential to accelerate computations to unprecedented speeds.
 The fundamental physics of optical signal transduction allows preserving the input-data timescales up to THz bandwidths, which is fundamentally out of range for electronics which usually are limited to GHz input data bandwidths. 
 However, mapping such concepts, in particular traditional ANNs and deep ANN architectures onto physical hardware induces significant challenges.
 These challenges are particularly pronounced in optical systems, where iterative and non-local training methods such as backpropagation are difficult to implement in hardware \cite{momeni2024training}, which leads to either using computationally expensive physical-twin optimization  \cite{wright2022deep} or gradient-free methods \cite{porte2021complete,skalli2024annealing,Nakajima2022}.

To significantly mitigate these challenges, single-layer feed-forward networks such as Extreme Learning Machines (ELMs) \cite{huang2004extreme} have been proposed as a hardware-friendly alternative \cite{ortin2015unified}.
 Extreme Learning Machines bypass the need for iterative weight optimization by maintaining the often randomly initialized input weights and biases fixed during training, while output weights can be computed in a single step using linear regression \cite{cucchi2022hands}.
 Conceptually speaking, ELMs transform input data into a high-dimensional representation by random nonlinear mappings, where many problems become linearly separable, simplifying computation without compromising accuracy.

Optical systems, including electro-optical \cite{ortin2015unified}, nanophotonics \cite{el2022photonic}, metasurfaces \cite{wu2021programmable}, and silicon photonics \cite{shen2017deep}, have shown particular promise as platforms for implementing ELMs.
 The high-speed nature of optical signal transduction enables implementing an ANN's connections without the usual limitations by a resistance and capacity time constant of charging an electronic circuit's transmission linear to a certain voltage.
 These connections are a key component of neural networks, and optics allows them to be efficiently performed in parallel, in some cases, enabling computation at the speed of light.

 It is important to emphasize that not only is linear mixing essential, but also nonlinear operations must be carried out in real-time and at ultrafast speed.
 In this context, the nonlinear propagation of short pulses in optical fibers arising from the instantaneous Kerr effect can potentially fulfill this requirement.
Recently, nonlinear fiber systems have been successfully exploited as physical reservoirs for pattern decoding and image classification within an ELM framework \cite{fischer2023neuromorphic}.

Here, we experimentally characterize a single-mode nonlinear fiber ELM’s principle computing metrics, such as effective dimensionality and consistency for future ultrafast timescale input data for the first time.
 Specifically, we use fs pulses injected into a highly nonlinear fiber to demonstrate an ELM system that can operate in real-time on ultrashort time scales. We encode information by modulating the spectral phase of short pulses injected into a highly nonlinear fiber.
 The ELM's internal layer state corresponds to the nonlinearly broadened optical spectrum at the fiber output, and the readout weights are digitally multiplied onto the spectrum to create the system's output in an offline procedure \cite{cucchi2022hands}.
 The ELM's nonlinearity is provided by the instantaneous optical Kerr effect, which can be described by the generalized nonlinear Schr\"odinger equation (GNLSE).
 Propagation induces continuous evolution of the nonlinear transformation, yet consecutive input states are disjoint due to the pulsed optical injection. 
  The ONN-topology (Optical Neural Network) is therefore a hybrid system as it inherits the internal nonlinear mixing from reservoir computing \cite{jaeger2004harnessing}, where each future state depends on the previous one, similar to time series. It is akin to the numerical process implemented when using the split step Fourier method based on the generalized nonlinear Schrödinger equation. However, our hardware also adopts the single-shot linear readout training from ELMs \cite{huang2004extreme}, which enables efficient training, yet does not allow for processing dynamical information, which in our case would be input-encoded onto subsequent input pulses. As such, nonlinear mixing in the system is akin to an RC (Reservoir Computing), yet the system does not exhibit working memory, which is akin to an ELM or a reservoir in its steady state.
 Our study focuses on understanding how fiber properties, such as dispersion, length, and nonlinearity, affect the system's computational capacity, which we gauge through task-independent evaluations \cite{skalli2022computational}.
 We experimentally characterize a physical ANN's principle computing metrics, effective dimensionality as well as consistency, as a function of input power for numerous fibers with different dispersion and nonlinear characteristics, as well as the dimensionality of injected data for the first time in an NLSE system.
 Finally, we assess the system's performance in classification tasks using the MNIST handwritten digit dataset for numerous combinations of input dimensionality and input power, also comparing the performance of an HNLF and a standard single-mode fiber.

\vspace*{-13pt}

\section{Concepts and optical computing hardware} 
\vspace*{-4pt}

\subsection{The structure of an Extreme Learning Machine}

RC and ELMs provide an efficient approach to ANN training by eliminating the iterative optimization of hidden layer parameters.
 As our HNLF optical computer lacks the ability of transient computing with the associated short-term memory, we will refer to our systems as an ELM.
 The network's architecture consists of an input layer comprising $d$ dimensions (neurons or features), a single hidden layer with $l$ dimensions, and an output layer with $k$ dimensions, see Fig.~\ref{fig:schematicsetup2}~(a).
 Considering input data with $N$ examples, the input data matrix $\mathbf{X} \in \mathbb{R}^{N \times d}$ represents the complete input data set.
 The distinguishing feature of ELMs lies in input weights and hidden layer biases remaining fixed throughout the training process.
 Often, they are even initialized randomly, yet in a hardware implementation, they are usually determined by the physics inherent to data injection as well as the nonlinear system's properties.
 The hidden layer performs a nonlinear transformation of the input data, and mathematically, this transformation is of the type 
\begin{equation}\label{eq:ELMstate}
\mathbf{H} = g(\mathbf{X} \mathbf{W}^{\text{in}} + \mathbf{b}), 
\end{equation} 
\noindent where $\mathbf{H}$ is the ELM's hidden layer state for all $N$ input examples, $\mathbf{W}^{\text{in}} \in \mathbb{R}^{d \times l}$ is a fixed and often randomly initialized matrix connecting input and hidden layer, $\mathbf{b} \in \mathbb{R}^{l}$ is the hidden layer bias vector, and $g(\cdot)$ is the ELM's nonlinear activation function.
 Common choices for $g(\cdot)$ in software ELMs include the sigmoid, ReLU, or hyperbolic tangent functions, and nonlinearity is required in order to enhance the feature representation of the injected data. Equation (\ref{eq:ELMstate}) represents the standard ELM formulation, but in our implementation, we do not have a bias term, (i.e., $b= 0$).
 In our experiment, $g$ is implemented through nonlinear propagation in the fiber, and $\mathbf{H}$ is the spectrum at the fiber's output.

The ELM's output $\mathbf{Y}$ containing $m$ features or neurons is computed for all $N$ input samples contained in $\mathbf{X}$ by linearly combining the hidden layer states according to 
\begin{equation}\label{eq:ELMout}
\mathbf{Y} = \mathbf{H} \mathbf{W}^{\text{out}}, 
\end{equation} 
\noindent where $\mathbf{W}^{\text{out}} \in \mathbb{R}^{l \times m}$ is the output weight matrix connecting the hidden layer to the output layer.
 In our experiment, $\mathbf{H}$ is determined with an optical spectrum analyzer (OSA), and weights $\mathbf{W}^{\text{out}}$ are digitally applied in a post-processing step.
 Training the ELM involves determining $\mathbf{W}^{\text{out}}$ such that the predicted outputs $\mathbf{Y}$ closely match the target outputs $\mathbf{T} \in \mathbb{R}^{N \times m}$.
 The objective is typically formulated as the minimization of a loss function, which often is the mean squared error between $\mathbf{Y}$ and $\mathbf{T}$: 
\begin{equation} \text{Loss} = | \mathbf{H} \mathbf{W}^{\text{out}} - \mathbf{T} |^2.
\end{equation} 
\noindent To solve for $\mathbf{W}^{\text{out}}$, ELMs leverage a closed-form solution through least-squares regression: 
\begin{equation}\label{eq:ELMwout} 
\mathbf{W}^{\text{out}} = \mathbf{H}^{\dagger} \mathbf{T}, 
\end{equation}
\noindent where $\mathbf{H}^{\dagger}$ is the Moore-Penrose pseudoinverse of $\mathbf{H}$.
 In order to avoid overfitting one often uses Ridge regression instead of Eq.~(\ref{eq:ELMwout}), however, when implementing the ELM in physical systems the ELM's internal as well as data-injection noise effectively regularizes $\mathbf{W}^{\text{out}}$ and overfitting is rarely a problem \cite{cucchi2022hands}.

This particular training concept has several key benefits for proof-of-concept implementations in hardware as well as for full hardware implementations, including in-situ learning \cite{porte2021complete,skalli2024annealing}.
 It bypasses the need for backpropagation and iterative optimization.
 Firstly, this enables rapid training by obtaining a final $\mathbf{W}^{\text{out}}$ in a single shot through standard numerical approximation methods like the \emph{pinv} function of Matlab \cite{MATLAB2024}.
 Secondly, training does not modify any parameters but the readout weights, hence, one can inject all $N$ input examples in one sequence and utilize a single recording of $\mathbf{H}$ for the entire training step.
 This would be impossible beyond ELM and RC architectures, where training inherently alters $\mathbf{H}$, which hence has to be re-determined has to be repeatably after each weight update.

\vspace*{-13pt}
\subsection{Experimental implementation}\label{sec:Experiment}
\begin{figure*}[tbp]
    \centering
    \includegraphics[width=\textwidth]{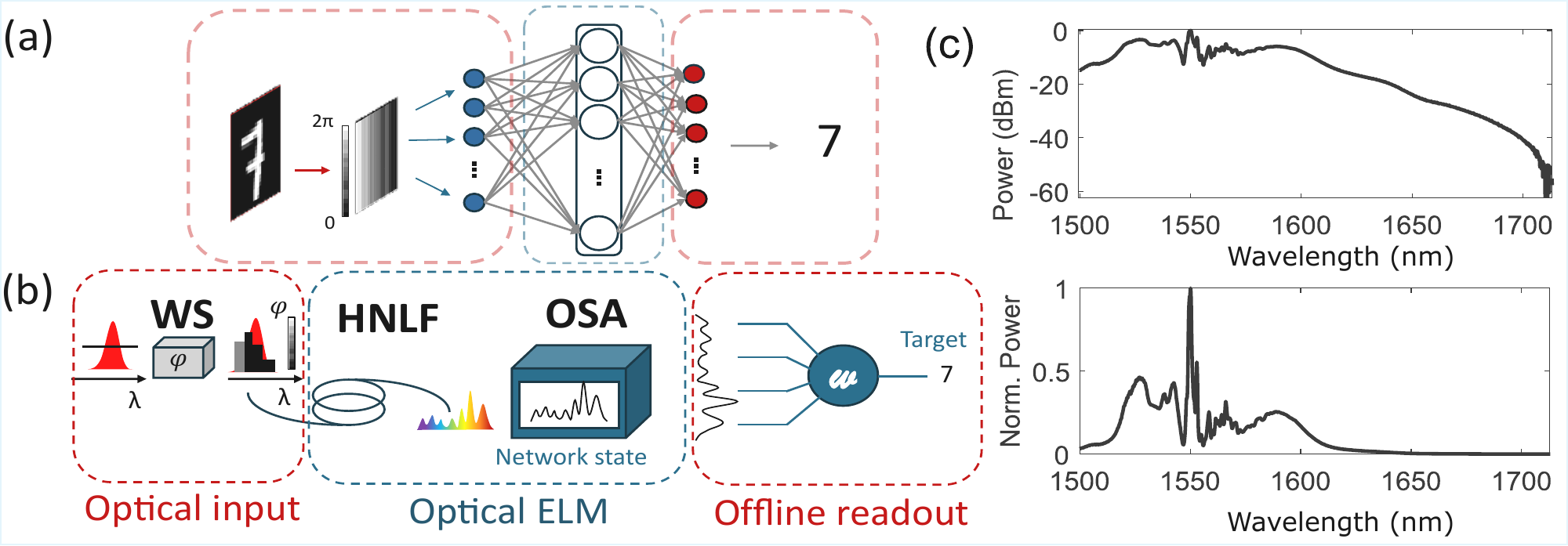}
    \caption{(a) conceptual diagram of the ELM (b) schematic of the experimental setup for the Extreme Learning Machine implementation composed of a waveshaper (WS), highly nonlinear fiber (HNLF), and an optical spectrum analyzer (OSA) to collect the output fiber spectra. Typical encoding patterns are depicted for both cases of task-dependent and task-independent. (c) Typical fiber output spectra in logarithmic and linear scale for an input dimension $d$ = 22 and an input power $P_{in}$ = 35~mW.}
    \label{fig:schematicsetup2}
\end{figure*}
The experimental setup, schematically illustrated in Fig.~\ref{fig:schematicsetup2}~(a),  used in this study builds upon a previously reported configuration \cite{hary2023tailored}. The pump laser (NKT Photonics ORIGAMI) delivers pulses with a duration of 235 fs at 1559.4 nm, a repetition rate of 40.9 MHz, a FWHM bandwidth of 14 nm, and a beam diameter of 5 mm.
 The setup features a 4-f line with a spatial light modulator (SLM, Holoeye Pluto 2.1) positioned in the Fourier plane in a folded arrangement. The incident beam is diffracted by a grating (G1) (600 lines/mm) in Littrow configuration. The horizontally dispersed beam is collimated by a lens (L1), (f = 30  cm), illuminates the SLM and propagates back the same path to be collected by the highly nonlinear fiber. The spot size of a single wavelength on the SLM, determined by the imaging resolution of the system, is approximately 122 $\mu$m, corresponding to around 15 pixels. Under such conditions, phase cross-talk between individual pixels is negligible. For simplicity, the figure depicts a transmissive SLM instead of reflective.

\begin{figure}[h]
    \centering
    \includegraphics[width=\textwidth]{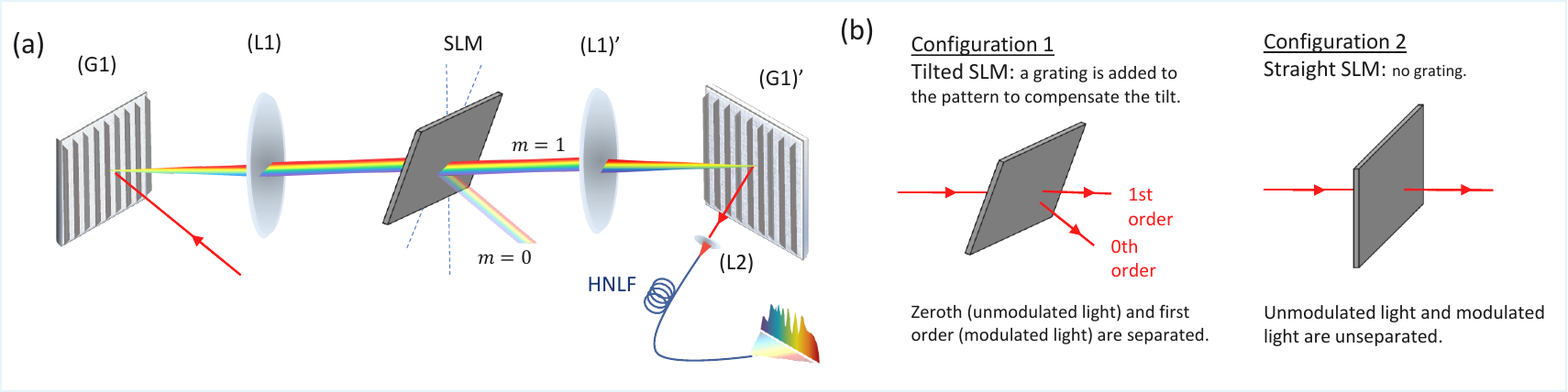}
    \caption{(a) conceptual diagram of the waveshaper used for data encoding, (b) schematic of the experimental SLM configurations. In Configuration 1, the SLM is tilted and a grating is applied to separate the orders spatially, allowing only modulated light to be coupled forward. In Configuration 2, the SLM is aligned straight, and no grating is applied, resulting in both modulated and unmodulated light propagating together.}
    \label{fig:schematic_config}
\end{figure}

Figure~\ref{fig:schematic_config}~(b) shows the two possible configurations that the SLM can be in, i.e., titled or straight. When the SLM is tilted, a vertical grating is added to the phase of the SLM, which compensates for the SLM tilt angle for its first diffraction order, such that only the first diffraction order is coupled into the fiber. A vertical blazed grating is added to each input pattern according to $X_{\text{SLM}} = X_{\text{G}} + \mathbf{u} \cdot \mathbf{W}^{\text{in}}$, $X_{\text{G}}$ is grating with a period of 16 pixels  (see Eq.~\ref{eq:ELMstate}), ensuring that only the first diffraction order is coupled into the fiber. The tilted configuration, therefore, suppresses the fraction of light that remains unmodulated by the spectral phase patterns, whereas the orthogonal configuration is easier to implement and merges modulated as well as unmodulated contributions in the same fiber. We assessed whether the inclusion of the unmodulated input pulse contributes additional computational dimensions by implementing both experimental conditions.
 For an input power of 39 mW, the generated spectral broadening spans from 1200 nm to 2200 nm when no phase pattern is applied.
 However, here the focus is on the 1500-2200 nm band to match the measurement range of the optical spectrum analyzer (OSA, Yokogawa AQ6376).
 The SLM is used to update the phase pattern of the input light, which is horizontally spread over 1000 SLM pixels, corresponding to the spread of the Gaussian pulse on the SLM, and repeated vertically 1080 times to match the height of the SLM.  The pixel size is 8~$\mu$m, and 14~nm input pulse spectrum covers $\sim$1100 pixels, resulting in $\sim$0.013~nm/pixel.
After propagation into the nonlinear fiber, the output spectrum is sampled across $l = 3000$ points spanning the wavelength range of 1500 to 2200 nm.
 In order to study the impact of nonlinear propagation in different dispersion regimes, we conducted experiments using multiple fibers with varying dispersion characteristics (see below for details).
 Each fiber was 5 meters long with 20 cm of standard single-mode fiber (SMF28) patch cords spliced in both ends, Fig.~\ref{fig:schematicsetup2}~(c) shows a typical output spectrum both in linear and logarithmic scale.

The maximal optical collection efficiency of the injected pulse was 55 $\%$.
 This was primarily limited by splice losses as well as the coupling efficiency from free-space propagation into the first SMF28 fiber.
 However, the spectral phase applied via the SLM results in diffraction, while injection into the SMF28 creates spatial filtering through the fiber's mode profile. 
 This can make the coupling efficiency sensitive to the input dimensionality $d$ and the particular phase pattern $\mathbf{X}$ displayed on the SLM.
 We systematically evaluate these effects in Section \ref{sec:TaskIndep}.

Three different HNLFs are used in our experiments, and their dispersion characteristics, over the range of 1500 to 1620 nm, are shown in Fig. \ref{fig:1_fiber_dispersion}~(a).
 At the pump wavelength $\lambda_{\text{pump}} = 1559.4$ nm, the dispersion coefficients are
\begin{itemize}
    \item $D_{\text{Fiber 1}} = -0.98$ ps/(nm·km) (normal dispersion, relatively flat phase profile).
    \item $D_{\text{Fiber 2}} = 0.046$ ps/(nm·km) (near zero-dispersion: anomalous dispersion above $\lambda_{\text{pump}}$ and normal dispersion below).
    \item $D_{\text{Fiber 3}} = 0.33$ ps/(nm·km).
\end{itemize}
 We used Fiber 1 with normal dispersion across the entire spectrum for each of the following evaluations, while in the section where we focus specifically on the impact of dispersion, we compare all three fibers.
 
\vspace*{-13pt}
\subsection{Principal Component Analysis}\label{sec:PCA}

Principal Component Analysis (PCA) is a linear statistical technique widely used for dimensionality reduction, feature extraction, and data visualization.
 By transforming the original data into a new set of uncorrelated variables, known as principal components (PCs), PCA captures the directions of maximum variance within a dataset.
 This identifies dominant patterns and allows mitigating redundancies while preserving as much relevant information as intended by keeping dominant PCs and discarding those below an ad-hoc defined relevance threshold.

To formally define PCA, let $\mathbf{X} \in \mathbb{R}^{N \times f}$ represent a dataset consisting of $N$ samples, each described by $f$ features.
 In this representation, each of the $N$ rows of $\mathbf{X}$ corresponds to a data sample, and each of the $f$ columns represents a feature.
 The goal of PCA is to project data $\mathbf{X}$ onto a set of orthogonal axes such that the variance along each axis is maximized.
 These axes are the PCs, and they are linear combinations of the original features $f$.

The PCA process begins with data centering, where the mean of each feature is subtracted to ensure that the dataset has a mean of zero, which is crucial for correctly computing the covariance.
 The centered dataset $\mathbf{X}^{\text{c}}$ is calculated as
\begin{equation}
\mathbf{X}^{\text{c}} = \mathbf{X} - \mathbf{1} \mathbf{\mu}^{(T)},
\end{equation}
\noindent where $\mathbf{1}$ is a $N \times 1$ column vector of ones, $\mathbf{\mu} \in \mathbb{R}^f$ is the vector of feature means and $(T)$ is the transpose of a vetor or matrix.
 Next, the covariance matrix $\mathbf{C} \in \mathbb{R}^{f \times f}$ is computed to capture the relationships between features by 
\begin{equation}
\mathbf{C} = \frac{1}{N-1} \mathbf{X}^{\text{c(T)}} \mathbf{X}^{\text{c}}.
\end{equation}
\noindent $\mathbf{C}$ encapsulates the variance of each feature along its diagonal and the pairwise covariances between features in its off-diagonal elements.
 The PCs are then derived from the eigenvalues and eigenvectors of the covariance matrix.
 Let $\mathbf{\Lambda} \in \mathbb{R}^{f \times f}$ be the diagonal matrix of eigenvalues and $\mathbf{V} \in \mathbb{R}^{f \times f}$ be the matrix of corresponding eigenvectors, such that
 \begin{equation}
 \label{eq: cvv}
 \mathbf{C} \mathbf{V} = \mathbf{V} \mathbf{\Lambda}.
 \end{equation}
\noindent The columns of $\mathbf{V}$ are eigenvectors providing the directions of maximum variance in the dataset, while the associated eigenvalues in $\mathbf{\Lambda}$ quantify the variance along each of these PCs.
 Sorting the diagonal entries in $\mathbf{\Lambda}$ in descending magnitude sorts the eigenvalues according to their relevance for the linear reconstruction of the original data $\mathbf{X}$.
 By selecting the top $k$ eigenvectors associated with the largest eigenvalues enables dimensionality reduction while preserving the most significant patterns in data $\mathbf{X}$.

The compressed dimensionality dataset, denoted as $\mathbf{X}^{\text{PCA}}$, is obtained by projecting the centered data onto the subspace spanned by the selected eigenvectors according to
\begin{equation}
\mathbf{X}^{\text{PCA}} = \mathbf{X}^{\text{c}} \mathbf{V}^{\text{PCA}},
\end{equation}
\noindent where $\mathbf{V}^{\text{PCA}} \in \mathbb{R}^{f \times k}$ contains only the top $k$ eigenvectors.
 This transformation reduces the dimensionality of the data from $f$ to $k$.
 
We employ PCA in two different settings throughout this work.

(i) In our task-independent analysis, PCA is used to determine the effective dimensionality of the optical ELM, serving as a tool to characterize the number of linearly independent spectral features generated by the system.

(ii) In the task-dependent analysis, PCA is used as a linear input compression method to investigate how varying the number of input features $d$ influences classification performance and dimensionality, see explanation in Section~\ref{sec:DimTuningMNIST}.

\noindent In this second case, the linearity of PCA aligns naturally with the linear readout structure of the ELMs at the output. While nonlinear methods may capture more structure, they would not be usable by the system's linear readout layer and thus would not improve task performance.

\section{Results}

In the following section, we investigate the ELM’s performance using various metrics.

\begin{figure*}[htbp]
    \centering
    \includegraphics[width=1\textwidth]{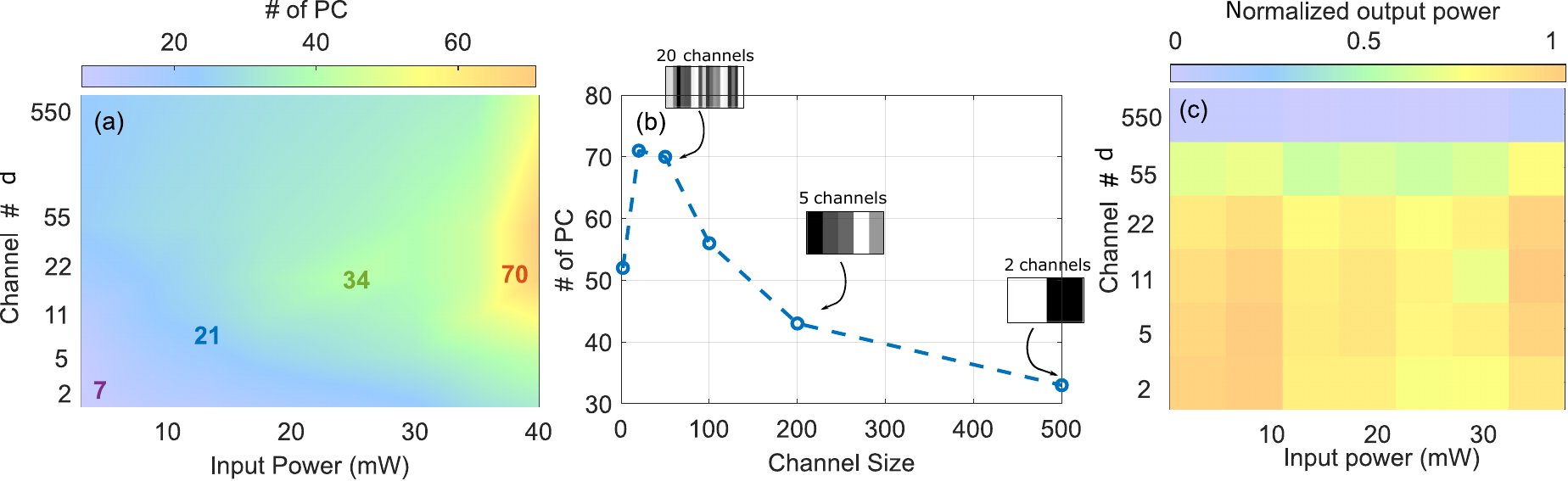}
    \caption{\textbf{(a)} Principal Component Analysis (PCA) results showing the effect of varying input power and channel size on the number of principal components (PCs. The color gradient indicates the number of PCs, with colder colors representing fewer PCs and cooler colors representing a higher number of PCs. This highlights the regions of channel size and input power combinations that maximize or minimize the complexity of the system response. \textbf{(b)} Number of PCs as a function of channel size for a fixed input power of 40 mW, illustrating the trend and identifying the optimal channel size for this input power level. The insets display example Spatial Light Modulator (SLM) patterns and their respective number of channels for selected channel sizes (50, 200, and 500), providing a visual representation of the corresponding SLM configuration at these sizes.
    \textbf{(c)} Integral over power spectral density measured by the OSA, normalized by the injected power.
    Optical losses are essentially independent of the injected power.}
    \label{fig:3_channel}
\end{figure*}

\subsection{Task-Independent performance metrics}\label{sec:TaskIndep}

In the context of ANN computing, the number of nonlinear (hidden) layers as well as the neurons within them are of determining importance.
 Final computing performance sensitively depends on these, and as such, both numbers are what determines the overall architecture of standard multi-layer perceptrons.
 Yet, in the context of unconventional computing substrates (i.e., non-digital and non-discrete elements), one can face a dilemma when attempting to specify the number of hidden layer neurons, here $l$.
 The number of hidden layer neurons $l$ is clearly defined when implementing an ELM in a classical software/code setting, or in hardware comprising discrete units acting as its neurons.
 However, when encoding ELM state $\mathbf{H}$ in a single high-dimensional and continuous physical state, then this dimensionality is obscured within a continuous spectrum of values \cite{skalli2022computational}, here the power spectral density of the broadened optical output spectrum obtained by the OSA.
 
To estimate the number of computational features, i.e., neurons in the output spectra, we apply PCA to finely sampled spectral data. This approach projects the spectrum onto a set of orthogonal dimensions, defined by the projection matrix $\mathbf{V}$ in Eq.(\ref{eq: cvv}), sorted by their contribution to the overall variance. While there is no strict equivalence between principal components and neurons, the number of significant PCs provides a heuristic estimate of the system's degree of freedom.

\noindent PCA is particularly suited to this task because it does not require a pre-defined functional basis, unlike other projection methods such as the one used in~\cite{dambre2012information}. This makes PCA more practical for both output analysis and input compression. Moreover, prior work has shown that orthogonality (and by extension dimensionality) \textcolor{black}{improve classification and regression performance \cite{LiOrthogonal21}. Such orthogonality} is maximized when the eigenvalue spectrum of connectivity matrices is flat (i.e., that most eigenvalues are roughly equal) \cite{LiOrthogonal21}, a property that supports our use of PCA as a dimensionality probe.
 
Although we did not apply PCA directly to the phase-encoded field before the fiber injection, we interpret the low power limit of our system, represented in Fig. \ref{fig:3_channel}~(c), as a proxy for the linear propagation case, where fiber nonlinearity is negligible. In this regime, the dimensionality is dominated by the SLM encoding and the intensity detection by the OSA. This provides a useful baseline, though we acknowledge that coherent phase encoding can introduce nontrivial structure, as shown in \cite{Yildirim2024}.
 If PCA would be applied directly to the input data, i.e. matrice of random independent phase patterns, then the number of principal components needed to explain the variance would be equal to the input dimensionality \( d \): all components contribute equally, and the explained variance is uniformly distributed.
 
 A femtosecond input pulse with $14$~nm bandwidth, whose spectral phase is modulated by the input data, can undergo significant broadening spanning up to hundreds of nanometers, through nonlinear propagation.
 However, this broadened spectrum does not necessarily imply high computational dimensionality. If the resulting spectral features are broad and respond in a similar way to changes in the input, then the system behaves as if it had only a few independent channels.
 In other words, it is not the spectral bandwidth of a spectrum but the number of linearly independent spectral features that determines its performance for computing. It is therefore not sufficient to only aim for maximal spectral broadening in the context of nonlinear fiber propagation for computing.

 In our experiment, we generate $N = 3000$ random patterns, each comprising $d$ random values uniformly distributed between 0 and $2\pi$, and apply PCA to the output spectra to evaluate the system’s effective dimensionality, denoted as $\text{PC}^{95}$, the number of PCs needed to explain 95\% of the spectral variance.
 Furthermore, we used matrix $\mathbf{H}$ to evaluate the system's consistency, which reflects its ability to produce stable and repeatable outputs, an essential property for reliable optical computing.

\subsubsection*{ELM dimensionality and physical input channel size}

\begin{figure*}[htbp]
    \centering
    \includegraphics[width=\textwidth]{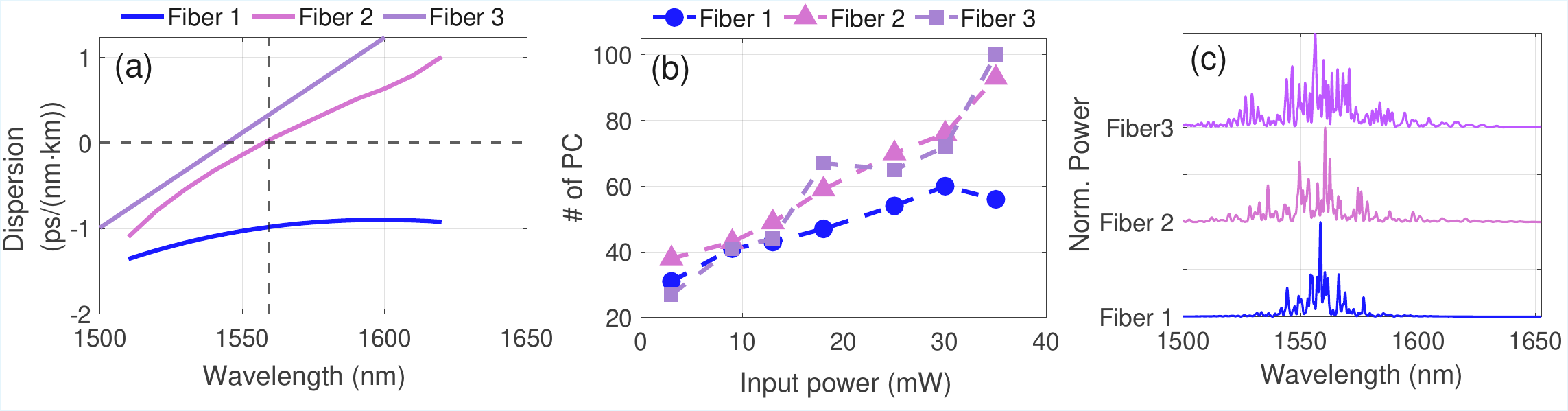}
    \caption{\textbf{(a)} Dispersion as a function of wavelength for three fibers (Fiber 1, Fiber 2, and Fiber 3). Fiber 1 exhibits normal dispersion with a relatively flat profile, while Fibers 2 and 3 demonstrate anomalous dispersion above the pump wavelength. \textbf{(b)} Number of principal components (PCs) versus input power for the three fibers. At low power, the number of PCs is similar across all fibers, and at higher power levels, Fibers 2 and 3 generate more PCs, indicating increased spectral complexity due to higher dispersion.
    \textbf{c} Three example spectra for the highest injection power and the three different fibers, revealing the systematic difference in spectral complexity and broadening between the normal (Fiber 1) and anomalous (fibers 2 and 3) dispersion.} 
    \label{fig:1_fiber_dispersion}
\end{figure*}

First on the list of our general HNLF ELM characterization is the impact of the physical size of each input channel.
 \emph{Channel size} refers to the number of adjacent SLM pixels grouped to form a single uniform phase region, varying in the range of 2, 20, 50, 100, 200, and 500 SLM pixels. For a total of 1100 illuminated pixels, this yields a number of channels, $d\in \{550, 110, 55, 22, 11, 5, 2\}$, where:
\begin{equation}
d = \frac{\text{Active SLM Area}}{\text{Channel Size}}.
\label{eq:slm_dimensionality}
\end{equation}
\noindent In principle, $d$ could range from 1 to 1100, limited by the number of illuminated SLM pixels, but in practice, we investigate configuration down to a minimum of $d = 2$, corresponding to 500 pixel-wide channels.

 The results, presented in Fig. \ref{fig:3_channel}, reveal the strong impact of input channel size and hence $d$ on the ELM's effective dimensionality $\text{PC}^{95}$ under varying input power levels.
 Overall, the dimensionality of the system increases with input power for all channel configurations.
 Figure \ref{fig:3_channel}~(b) focuses on a fixed input power of $P_{\text{in}}=40~$mW, for which we obtain a maximum value of $\text{PC}^{95}=70$ for channel sizes between 20 and 50 pixels.
 Insets in Fig. \ref{fig:3_channel}~(b) illustrate typical SLM patterns corresponding to selected channel sizes (50, 200, and 500), providing a visual representation of the input configuration.
 These results suggest that intermediate channel sizes strike a balance between maintaining sufficient spatial resolution and preserving the diversity of spectral features, thus maximizing the computational capacity of the system.

Besides the effectivity of the underlying nonlinear dynamics for creating relevant ELM dimensionality, one also needs to consider the potential influence of linear effects.
 In Fig.~\ref{fig:3_channel}~(c), we show the integrated optical power spectral density (PSD) measured after propagation through the optical fibers and normalized by the input power as a function of $d$.
 Spectral phase modulations with a spatial period smaller than 20 SLM pixels result in a noticeable drop of optical power at the fiber output.
 Additionally, we examined this effect for a wide range of input powers, which reveals no systematic impact.
 This rules out nonlinear dynamic effects during propagation in the HNLF as underlying cause. 
 We therefore associated this drop of output power exclusively with a loss of input coupling efficiency.
 The application of spatial phase modulation of the input beam by the waveshaper results in the broadening of the focused image in the waveshaper's Fourier plane, where the pulse is injected into the SMF28 fiber.
 For a phase modulation with a sufficiently high spatial frequency, the input field at the fiber facet will increasingly deviate from an LP01 mode, and coupling efficiency will therefore drop as a consequence, which is what can be seen in Fig.~\ref{fig:3_channel}~(c).
 
\subsubsection*{ELM dimensionality and fiber dispersion}

Dispersion is one of the main influences determining the characteristics of propagation in NLSE systems, and such of major relevance in our context \cite{marcucci2020theory}.
 The characteristics of spectral broadening depend sensitively on the dispersion properties of the fiber used and the associated laser pump wavelength, as these determine the details of the underlying propagation dynamics.
 Specifically, when a spectral broadening is seeded in the normal dispersion regime using femtosecond or picosecond pulses, the spectral broadening is typically dominated by self-phase modulation and yields smoothly varying spectral features.
 In contrast, seeding in the anomalous dispersion regime leads to more complex dynamics such as soliton fission, dispersive wave generation, and the Raman self-frequency shift, generally resulting in significantly more fine structure on the output spectra as well as broader spectral widths (at the same input pulse energy).
 These differences and the following spectral-feature widths, as well as their linear correlation, can hence have a significant influence on the system's number of effective dimensionality for computing.

For this test, we fixed $d=50$ input dimensions.
 The influence of fiber dispersion on the number of PCs was evaluated for the three distinct fibers introduced in Section~\ref{sec:Experiment} exhibiting normal, near-zero, and anomalous dispersion regimes.
 The number of effective PCs was computed for input powers $P_{\text{in}} \in \{3, 9, 13, 18, 25, 30, 35\}$ mW, with the results shown in Fig. \ref{fig:1_fiber_dispersion}~(b).
 At low power levels ($P_{\text{in}} \leq 13$ mW), the number of PCs remained relatively consistent across all fibers.
 However, at higher powers, the low and normal dispersion Fiber 1 produced significantly fewer PCs compared to Fibers 2 and 3.
 The maximum number of PCs for Fiber 1 was $\sim60$ at $P_{\text{in}} = 30$ mW, while Fibers 2 and 3 with anomalous dispersion regime dynamics exhibited approximately $\text{PC}^{95}\approx100$ at the same $P_{\text{in}}$ value.
 This suggests that the more complex spectral broadening mechanisms associated with soliton dynamics lead to output spectra on which more linearly separable features can be encoded.
 We attribute this physically to the lower degree of internal correlation across a broadband spectrum when compared to spectra generated from the normal dispersion dynamics of self-phase modulation.
 The systematic differences between the normal (Fiber 1) and anomalous (Fibers 2 and 3) propagation on broadening and number of spectral features around the seeding wavelength can be seen in Fig.~\ref{fig:1_fiber_dispersion}~(c).
 Furthermore, the number of features around the pump for Fiber 1 is smaller than for fibers 2 and 3, explaining the difference in $\text{PC}^{95}$ and providing a first indicative confirmation of the earlier presented argument regarding the impact of dispersion.

\subsubsection*{Spectral location of computing dimensions}

\begin{figure*}[htbp]
    \centering
    \includegraphics[width=\textwidth]{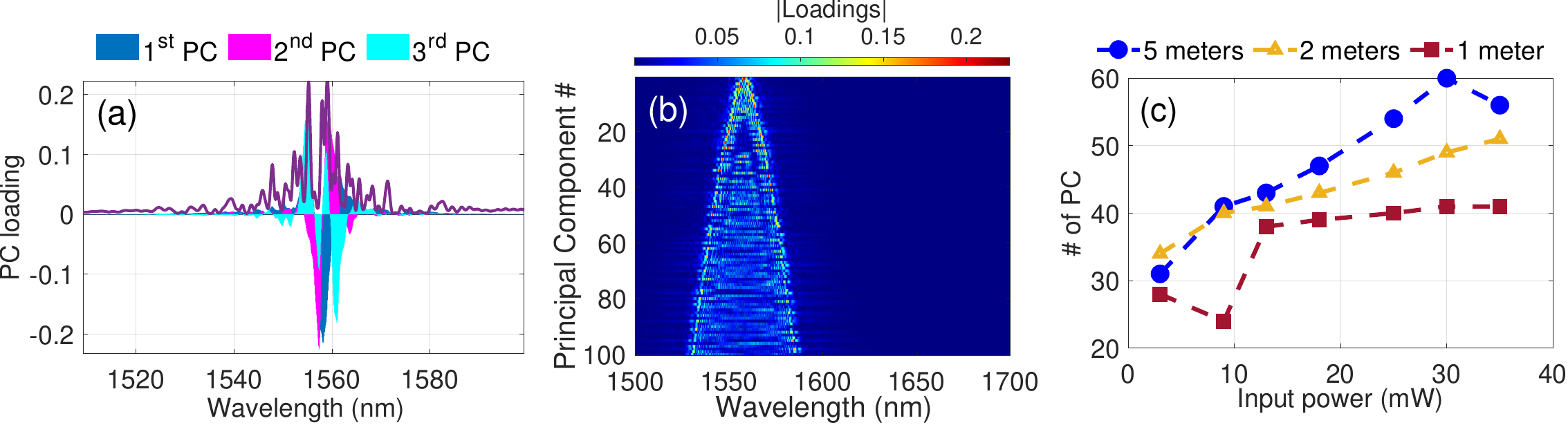}
    \caption{(a) Spectral distribution of the PCs shows that the ELM's feature space relies mostly on dynamics close to the pump wavelength.(b) Spectral loadings of the first 100 PCs as a function of wavelength for the case of 35 mW, Fiber 3.
    (c) Number of PCs as a function of input power for three different fiber lengths (1, 2, and 5 meters). The 5-meter fiber consistently exhibits a higher number of PCs compared to shorter fibers, indicating that longer fibers enhance the dimensionality of the output spectra.}
    \label{fig:2_length}
\end{figure*}

One of the striking features of nonlinear fiber propagation according to the NLSE is their capacity to transform an originally $\sim$14~nm broad input pulse into spanning several hundred nanometers.
 Encoding some 40 input features on these 14~nm spectral input bandwidth corresponding to $\sim2.6$~features/nm.
 Extrapolating this to the output spectrum spanning several hundred of nanometers would result in the order a thousand features, yet as shown in the previous section, this is not the case.
 This raises the question of where these spectrum's degrees of freedom are located, i.e., which part of the spectral broadening contributes significantly to the $\text{PC}^{95}$.
 Since PCA is a linear technique based on projecting the original features, here the sampling positions of the optical spectra, onto the orthogonal space of PCs via $\mathbf{V}$, one can use this matrix to show how individual PC eigenvectors are associated with spectral positions.
 Figure~\ref{fig:2_length}~(a) shows this spectral loading for the first three principal components for $P_{\text{in}}=35$~mW and $d=50$ input samples, combined with the original optical output spectrum on a linear scale.
 One can clearly see that the wavelengths involved in creating the $\text{PC}^{95}$ are located very closely around the input pulse, roughly within a width of $\pm40~$nm. This spectral localization is consistent with previous observations in similar nonlinear systems \cite{fischer2023neuromorphic}. The first three principal components account for less than 20 $\%$ of the total spectral loading, and approximately 13 components are required to reach more than 50$\%$. This raises the question of whether the spectral components outside the pump bandwidth contribute meaningfully to the remaining components.

\noindent To explore this, we computed the absolute value of the spectral loading distributions for the first 100 principal components as shown in Fig. \ref{fig:2_length}~(b). The dominant 10 components are tightly localized around the pump, while components in the range of 20-30 display spectral loadings that are shifted towards the wings. Even after the first 100 PCs, the spectral loading only approaches a width of $\sim100~$nm.

This has two main reasons, that are linked to the particularities of computing.
 Firstly, the PSD of spectral broadening dynamics spanning hundreds of nanometers is mostly discernible when employing a logarithmic dB scale.
 However, a computer output relies on combining these different spectral intensities according to some ratio. In order to leverage the different features encoded in wavelengths, one needs to be able to linearly balance these different features to create a desired output. To illustrate the consequence for computing of this operation, if one wants to leverage two independent spectral components that have 60 dB difference in their relative power,(corresponding to typical OSA dynamic range and using the full SC spectrum), one would first have to balance their power before their different contributions can be relatively scaled according to $\mathbf{W}^{\textrm{out}}$ to equally contribute to computing. As each independent spectral component would require such a normalization, transforming such a wide PSD range spectrum into components of equal relative PSD necessitates programmable weights with a resolution of $1/(60~\textrm{dB})=10^{-6}$, corresponding to $\log_{2}(10^{-6})=19.93$ or 20 bits. After that stage additional readout weights can be added in order to make the computer programmable, and assuming 8 bit resolution for $\mathbf{W}^{\textrm{out}}$ would result in the usually unattainable 28 bits overall required for implementing a sufficiently resolved and programmable optical computer with a PSD range spanning 60~dB.

Secondly, a simple visual analysis of typical spectra shows that around the pump wavelength, different spectral features are quite narrow, of the order 1~nm, while these widths change to 10s of nm in the regions of the spectral broadening further away from the pump.
 Linearly independent dynamics creating independent PCs can only be provided by spectral components that are also linearly uncorrelated as a response to input modulations, and the lower limit of spectral density of such uncorrelated components is determined by their spectral widths, similar to the Rayleigh criterion in imaging.
 Broadband soliton dynamics, for example, mainly responsible for spectral broadening, can therefore only contribute in a limited capacity to an optical ELM's computational dimensionality $\text{PC}^{95}$.
 
 We have measured these loadings for both waveshaper configurations (Fig.~\ref{fig:schematic_config}~(b)),  including/excluding the $0^{\text{th}}$ diffractive order in the injected signal, and no systematic difference was found.
 Heuristically speaking this is reasonable; compared to $1^{\text{st}}$ order only spectral broadening seeding, including the $0^{\text{th}}$ order creates an interference between the fully phase-modulated $1^{\text{st}}$ order and the un-modulated part of the $0^{\text{th}}$ order.
 Interference corresponds to the superposition of the two optical fields, and in a physics context, the superposition principle is inherently linear, hence does not augment the feature dimensionality, yet it could potentially be interpreted as influencing injection weights $\mathbf{W}^{\text{in}}$.
 While intensity detection introduces an interference term of the form $2\mathrm{Re}\{E_1 E_2^*\}$, $E_1$, $E_2$, the field corresponding to the zeroth and first diffraction order respectively. This term depends only on the relative amplitude and phase between the two interfering fields and does not scale with the overall power. Therefore, it does not act as a power-dependent nonlinearity like the one that arises from the fiber, via the Kerr effect. The interpretation is consistent with our experimental findings that no increase in the dimensionality was observed when the light is not separated into diffraction orders.
 For $1^{\text{st}}$ order only spectral broadening seeding we added a constant blazed grating to the phase profile of the SLM and only collected the $1^{\text{st}}$ diffracted order, which was achieved by adding a grating pattern to $\mathbf{X}$.

\subsubsection*{ELM dimensionality and fiber length}

We investigated the impact of fiber length as another critical parameter influencing spectral broadening and hence potential computational capacity.
 The experiment was repeated for Fiber 1 at lengths of 1, 2, and 5 meters. The number of PCs as a function of input power is shown in Fig. \ref{fig:2_length}~(c).
 The results reveal a clear dependence on fiber length.
 The 5-meter fiber consistently produced a higher number of PCs compared to the shorter fibers at equivalent input powers.
 For instance, $\text{PC}^{95}$ increased from approximately 30 to 60 for the 5-meter fiber as $P_{\text{in}}$ increased from 3 to 35 mW.
 In comparison, the 1-meter fiber exhibited a slower increase, with $\text{PC}^{95}$ rising from approximately 30 to 40 over the same power range.
 These findings suggest that longer fibers enhance the dimensionality of the output spectra.

\subsection*{Consistency of the ELM}\label{sec:Consistency}

\begin{figure*}[htbp]
    \centering
    \includegraphics[width=\textwidth]{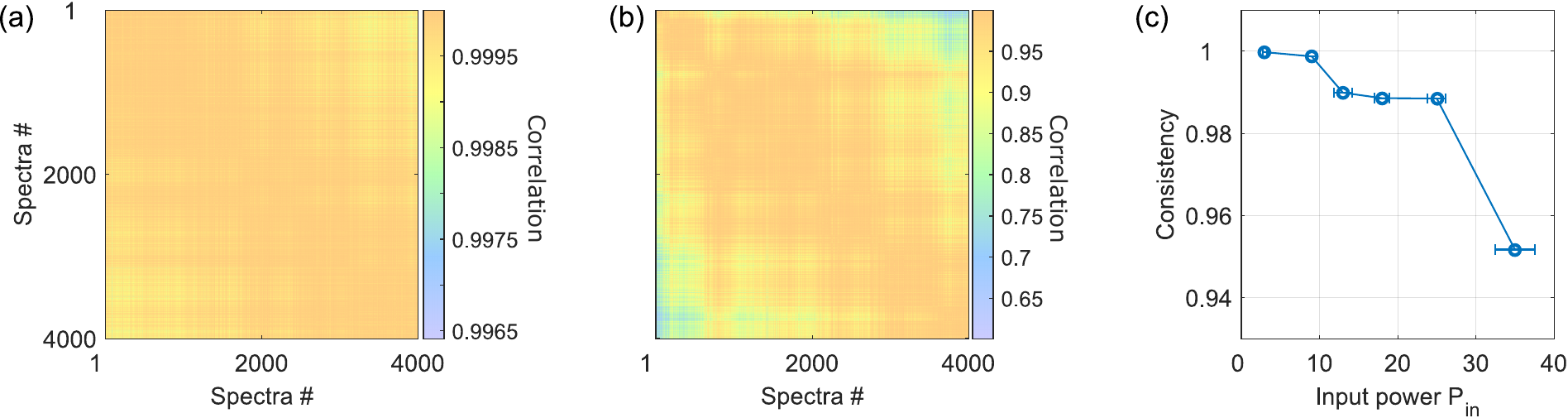}
    \caption{Consistency of the ELM's hidden layer state for 4000 repetitions of the same random input $\mathbf{X}$ with $d=50$.
    A computer's consistent reply to identical input information is a fundamental requirement for computing.
    (a) and (b) show the correlation matrix between the 4000 different spectra of $P_{\text{in}}=3$~mW and $P_{\text{in}}=35$~mW, respectively.
    For the low injection power, the reply is close to perfectly consistent, while for $P_{\text{in}}=35$~mW the correlation drops down to $\sim0.6$ at the end of the recording.
    (c) Consistency, corresponding to averaging the upper triangle of the correlation matrix.
    Consistency monotonously decreases with $P_{\text{in}}$, with a significant drop for $P_{\text{in}}>25$~mW.}
    \label{fig:Consistency}
\end{figure*}

The dimensionality described in the previous sections is, however, only one-half of the fundamental requirements for an ANN-based function approximator such as our physical ELM.
 The other half is consistency, which determines the capacity of such a system to respond in a reliable and reproducible manner to identical input information.
 A heuristic analogy can be made to a classical computer program when ignoring exotic examples of chaotic system simulations and comparable scenarios: executing the same computer code numerous times using the same user or data input generally produces the same outcome with a very high probability.
 The opposite, the same code leading to different outcomes, usually renders the computer or the code incapable of addressing relevant computing tasks.
 In dynamical systems, the capacity to react to repeated input with similar responses is determined by a system's consistency \cite{uchida2004consistency,lymburn2020quantifying}, and inconsistency renders reproducible computing with fully chaotic systems impossible until today.
 For that, we constructed $\mathbf{X}$ using $N=4000$ and $d=50$, where the $N$ input examples are all identical, consisting of a random and uniformly distributed 50 phase values.
 The concatenated system's output spectra $\mathbf{H}$ are then used to calculate the correlation coefficient between all $N$ responses, and the resulting correlation matrix is shown in Fig.~\ref{fig:Consistency}~(a,b) for $P_{\text{in}}=3$~mW and $P_{\text{in}}=35$~mW, respectively.

Each of the $N$ responses in $\mathbf{H}$ is an optical spectrum averaging numerous pulses, and as such the signal-to-noise ratio can be very high, as can be seen by the above $99.7\%$ correlation in Fig.~\ref{fig:Consistency}~(a) that was obtained for $P_{\text{in}}=3$~mW.
 Importantly, such high correlations are also regularly achieved with real-time implementations.
 Two signatures are visible from these measurements.
 The first is the spectra-to-spectra fluctuations, and the other is the long-term drift of such a system.
 The impact of the first is visible as changes in correlation values between neighboring spectra, while the latter are located towards the top-right and bottom-left corners.
 The correlation matrix is symmetric to its diagonal since zero-lag correlations in non-quantum systems are symmetrical.
 However, when increasing the optical injection power to $P_{\text{in}}=35$~mW, the general level of correlation drops substantially, with a minimum of $\sim62\%$, see Fig.~\ref{fig:Consistency}~(b).
 Furthermore, the overall structure too is altered, now exhibiting significant drops in correlation between directly neighboring spectra.
 The system's stability is therefore reduced in the context of both spectra-to-spectra correlations as well as with regard to long-term stability.

A system's consistency is determined by the average value to the correlation matrice's upper triangle \cite{uchida2004consistency}, and in Fig.~\ref{fig:Consistency}~(c) we show the consistency's dependency for a range of $P_{\text{in}}$.
 Consistency monotonously reduces for increasing $P_{\text{in}}$ until it essentially collapses for $P_{\text{in}}>25$~mW.
 Such loss of consistency is regularly exhibited when computing with nonlinear dynamical systems \cite{marquez2018dynamical}.
 Here, the observed drop in the correlation at higher input power could be caused by a loss of spectral coherence, a commonly observed occurrence in nonlinear spectral broadening \cite{dudley2006supercontinuum}.  Importantly, here we have tracked the power-collection efficiency during the measurement, with their standard deviations given as $P_{\textrm{in}}$ error bars.

\subsection{Task-dependent performance metrics: MNIST Handwritten Digits Classification} 

\subsubsection*{MNIST dimensionality tuning via PCA}\label{sec:DimTuningMNIST}

The MNIST handwritten digit dataset was used as a benchmark task to evaluate the performance of the optical setup as a computing medium.
 The dataset consists of handwritten digits from 0 to 9, providing a widely used standard for classification tasks.
 In the previous section, we identified a non-monotonous relationship between input dimensionality $d$ and $\text{PC}^{95}$, most importantly showing a decline of $\text{PC}^{95}$ for $d>50$.
 However, the native MNIST handwritten digits datasets comprise images of $28\times28=748$ pixels, which, as demonstrated, will reduce the dimensionality of our ELM.
 Most importantly, following Cover's theorem, ELM computing leverages dimensionality expansion as a mechanism for improving computational performance as compared to simply using the input data connected to a linear layer.
 The input dimensionality $d$ used for encoding the MNIST digits is therefore of major relevance.
 Using PCA is one possibility to adjust $d$ while other approaches could alternatively utilize nonlinear interpolation concepts or simple down-sampling according to linear interpolation.
 However, linear down-sampling does not take the importance of certain regions into account, while nonlinear interpolation is challenging in optics to implement in hardware as it requires nonlinear spatially extended, i.e., multimode optical transformations that typically are power hungry.
 PCA is a linear method and hence can be implemented using linear optics \cite{ma2019photonic,de2016scalable}.
 
\begin{figure}[htbp]
    \centering
    \includegraphics[width=1\columnwidth]{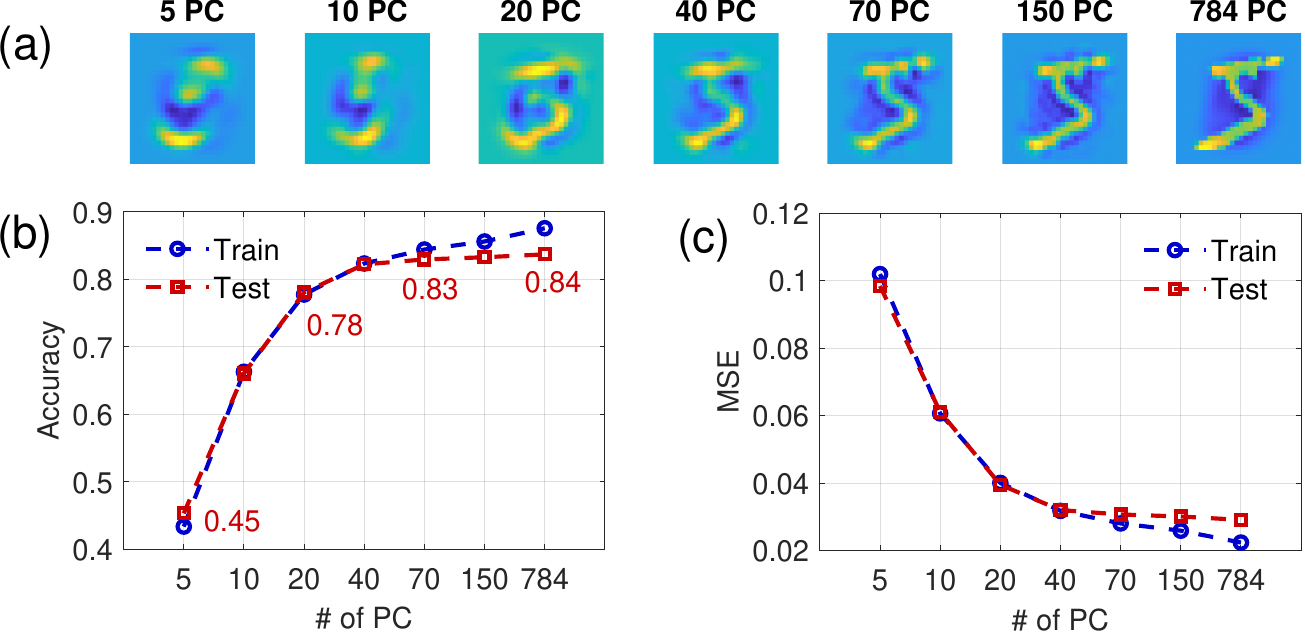}
     \caption{(a) Reconstructing a sample MNIST image using $k$ $\in$ [5, 10, 20, 40, 70, 150, 784] principal components (PCs), illustrating the gradual improvement in image fidelity as more PCs are used.
     Training as well as testing (a) accuracy and (b) MSE for a linear classifier as a function of the number of PCs, showing improved accuracy with increased PCs.
     These results represent the linear benchmark that the optical ELM needs to surpass in order to claim relevant computing.
     Testing performance saturates for $k>70$ PCs.
     Training is done on 10000 examples, while testing is done on 1000.}
    \label{fig:4_mnist}
\end{figure}

We apply PCA for image compression, as described in Section \ref{sec:PCA}, to adjust the number of pixels in which we encode the MNIST images \cite{lecun1998gradient}. This dimensionality reduction is performed on a classical computer before optical encoding.
 This reconstruction retains only the information captured by the top $k$ PCs, with fidelity increasing as $k$ grows; with $k = 784$, the reconstruction becomes exact.
 Figure \ref{fig:4_mnist}~(a) demonstrates the reconstruction of a sample image ('5') for a range of PCs.
 With only 5 PCs, the image is highly blurred and substantially lacks detail, by the eye it is hardly identifiable for a human.
 As the number of PCs increases to 20 and beyond, more features become discernible, the digit becomes progressively clearer and visibly identifiable as a '5'.
 At 150 PCs, the reconstructed image closely resembles the original, with minimal information loss.
 These examples highlight the relationship between the number of PCs and the retention of relevant information, demonstrating that only a subset of PCs is required to preserve the essential characteristics.

Figure \ref{fig:4_mnist}~(b) and \ref{fig:4_mnist}~(c) present the classification accuracy and MSE (mean square error) results for training and testing datasets, using varying numbers of PCs.
 For that the $d=k$ input channles were directly connected to the $m=10$ outputs through a linear matrix operation $\mathbf{Y}=\mathbf{X}^{(k)}\mathbf{W}^{\text{out}}$, where $\mathbf{X}^{(k)}$ is the set of $N$ MNIST digits compressed to $d=k$ input dimensions via PCA.
 The output weight matrix $\mathbf{W}^{\text{out}}$ was computed using the Moore-Penrose pseudoinverse method as per Eq.~(\ref{eq:ELMwout}), and the same weights were applied to both training and testing data.
 Training was done on 10,000 examples while using 1,000 examples for testing.
 This linear classification accuracy increases significantly with the number of PCs, particularly between 5 and 40 PCs, where the testing accuracy improves from 0.45 to 0.83.
 Beyond 40 PCs, test accuracy gains start diminishing, with training accuracy continuing to rise.
 This plateau suggests that higher PCs primarily add redundant or noisy information, contributing little to the generalization capability of linear classifier $\mathbf{W}^{\text{out}}$ for this size of the training data set.
 Similarly, the MSE decreases sharply between 5 and 40 PCs, with slower improvements thereafter.
 This linear classifier we set as our lower benchmark limit that our ELM needs to surpass if we want to claim the ELM is actually performing relevant computing.
 Below this threshold, it would be computationally beneficial to compute linearly directly on the input data, which is more efficient in the general sense.
 Importantly, this linear limit is dataset-dependent, as applying PCA to non-random input data, like MNIST, yields a dimensionality that reflects the information content of the dataset. This is precisely what we observe in our PCA analysis of MNIST inputs, as shown in Fig.~\ref{fig:4_mnist} (b), (c).

\subsubsection*{Results with the Highly Nonlinear Fiber}

\begin{figure*}[htbp!] \centering \includegraphics[width=1\textwidth]{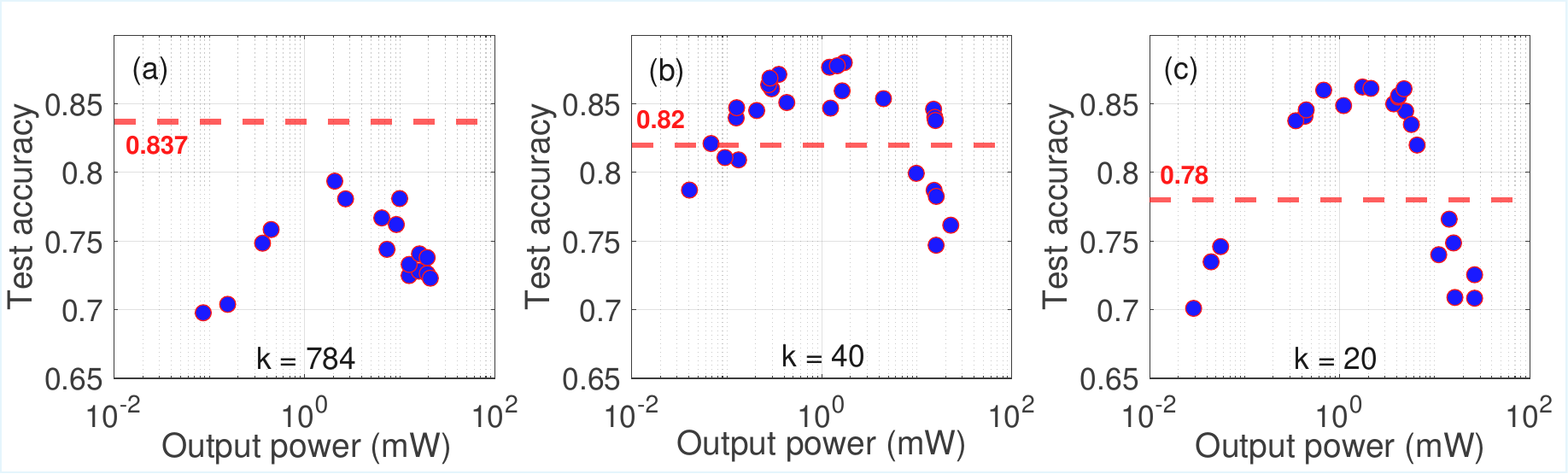} \caption{Test accuracy as a function of input power for different encoding dimensions: \textbf{(a)} 784 PCs, \textbf{(b)} 40 PCs, and \textbf{(c)} 20 PCs. The red dashed line represents the linear baseline accuracy, achieved without the nonlinear fiber effects. Higher encoding dimensions (784 PCs) show limited improvement in accuracy, while intermediate dimensions (40 PCs) exceed the baseline at most power levels. Lower dimensions (20 PCs) exhibit sensitivity to power, with peak performance within a specific power range.} \label{fig:5_ELMresults} 
\end{figure*}

For characterizing the optical ELM, the MNIST dataset was PCA encoded with $k=d \in \{20, 40, 784\}$. They are then encoded onto the SLM as phase masks, where each component is mapped to a grayscale level corresponding to a phase value.
 The modified optical pulse then propagated through HNLF Fiber 1.
 This experiment was conducted using 10,000 sample images, using a random selection of 80$\%$ of the resulting spectra for training and the remaining 20\% for testing.
 The system's training and classification accuracy on the test set was evaluated at various input power levels, as shown in Fig. \ref{fig:5_ELMresults}.
 There, the red dashed line provides the linear limit for each $k$ as explained in the previous section.

Figure \ref{fig:5_ELMresults}~(a) presents the results for the MNIST images with full resolution $d=784$.
 In this configuration, the system significantly falls below the linear baseline accuracy, with accuracy consistently below 0.837 at any input power level. Note that this linear baseline of 0.837 refers to the PCA-based dimensionality obtained by applying principal component analysis directly to the input source data (the first 10000 examples of MNIST), without propagation through the nonlinear fiber. This serves as a reference for the system and should not be confused with the performance of a linear classifier, for $d=k=784$.
 An optimal power level is observed around $P_{\text{in}}=2$~mW, achieving an accuracy of approximately 0.8. Beyond this point, no significant improvements are observed, indicating a saturation effect where higher power fails to enhance performance.
 Panel (b) shows the results for encoding with $d=k=40$.
 What is noteworthy is that here the system performs significantly better than when using the fully dimensional input with $d=k=784$.
 Importantly, across most input power levels, the test accuracy exceeds the linear baseline of 0.82, peaking at 0.880 at $P_{\text{in}}=1.72$~mW.
 However, at higher power levels, the accuracy drops to $\sim0.75$, suggesting that excessive power introduces non-ideal interactions in the HNLF.
 Panel (c) shows the results for $d=k=20$, where the system obtains a striking result, achieving nearly 10 $\%$ higher accuracy than the linear baseline. The performance is optimal within a specific power range of $P^{\text{in}}=0.4 \dots 14$~mW, and again accuracy decreases sharply outside this range.
 Best performances at the lowest $P_{\text{in}}$ for similar performance are summarized in Table~\ref{tab:comparison}.
  
These results provide very relevant insights for using NLSE substrates for physically implementing ELMs, here in the context of an HNLF.
 Two effects have to be considered with regard to the best performances obtained.
 First, as shown in Fig.~\ref{fig:3_channel}, for $d>50$ the optical ELM suffers from excessive injection losses, which is detrimental to the ELM dimensionality expansion.
 Second and more important are the considerations of input dimensionality $d$ and ELM dimensionality $\text{PCA}^{95}$ in the context of Cover's theorem.
 Generally, our ELM's dimensionality is limited to around $\text{PCA}^{95}=70$ for all tested configurations.
 This means by the fundamental properties of random dimensionality expansion that the optical ELM can aid computation only if $\text{PCA}^{95}>d$.
 For $\text{PCA}^{95}\leq d$ the optical system acts as a random dimensionality ``reducer", not an expander.
 Consequence of Cover's theorem, classification under such conditions becomes harder using random projections.
 These are necessary, not sufficient criteria.
 
Finally, the drop in performance for powers exceeding a certain threshold is related to our consistency measurements in Section~\ref{sec:Consistency}.
 Beyond this point, responses of the network become inconsistent \cite{uchida2004consistency} for repeated injections of identical input information.
 As such, the system's dimensionality potentially grows, which in principle helps the approximation property, yet the system loses its consistency property, and repeated injections of the same input will lead to different results. This trade-off aligns with previous observations in nonlinear photonics systems \cite{MudaTEgin25}. Our findings support the interpretation that there exists an optimal range of nonlinearity, sufficiently strong to enrich the feature space, but not so strong as to compromise functional stability. However, this threshold is inherently task-dependent, and it would likely vary if a different dataset or input structure were used.
 
\begin{table} [!h]
	\centering
	\caption{Performance comparison between linear and nonlinear systems for different encoding configurations.}
	\begin{tabular}{lrrr}
			System & 784 Channels & 20 PC & 40 PC \\
		\textbf{Linear (benchmark)} & 83.7\% & 78\% & 82\% \\
		\textbf{Nonlinear (experiment)} & 75.5\% & 86\% & 88\% \\
		\textbf{Minimum output power} & $\sim$2~mW & $\sim$0.5~mW & $\sim$0.7~mW \\ 
	\end{tabular}
	\label{tab:comparison}
	\vspace*{-0.5pt}
\end{table}

Table \ref{tab:comparison} summarizes the best performance results for the MNIST dataset.
 Other optical systems have demonstrated strong classification performance. For example, nonlinear multiple scattering cavities have been used to reach accuracies above 80\% on the Fashion-MNIST dataset \cite{Xia2024}. A multimode fiber system with spatial encoding achieved 94.5\% \cite{RahmaniTegin22}, while recent work based on Rayleigh scattering reports similar performance \cite{Redding2024}. Similarly, highly nonlinear fibers have been employed for optical ELMs with spiral encoding, with classification accuracy reaching 91\% \cite{saeed2025Chemnizt}. Diffractive free-space systems have shown promising results as well, when combined with nonlinear transformations by semiconductor lasers, achieving accuracies up to 96.5\%~\cite{skalli2025modelfreefronttoendtraininglarge} without pre- or post-processing, and integrated photonics approaches have achieved up to 97\% using on-chip tensor cores~\cite{Meng2025}. Similarly, spiking optical platforms~\cite{Robertson2022} have reported competitive performance.

\subsubsection*{Comparison Between HNLF and SMF}

\begin{figure}[htbp] 
\centering 
\includegraphics[width=0.5\textwidth]{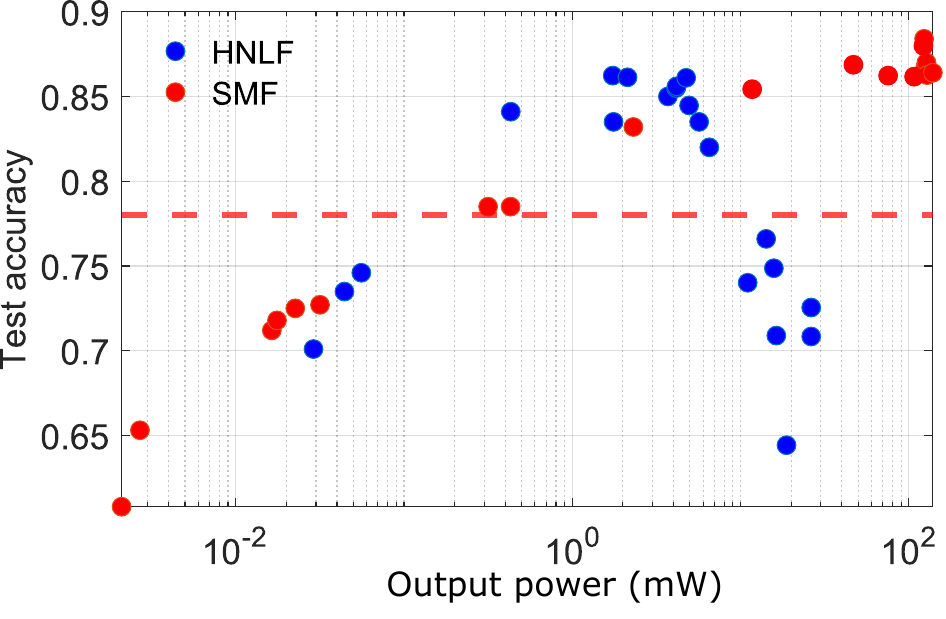} \caption{Comparison of test accuracy for HNLF (blue circles) and SMF (red crosses) as a function of input power. The HNLF, with a smaller core, exhibits nonlinear effects at lower power levels, while the SMF requires higher powers due to its larger core and longer nonlinear length. The red dashed line indicates the linear baseline accuracy.} \label{fig:6_elm_smf} \end{figure}

To further investigate the impact of fiber properties on system performance, a comparison was conducted between the HNLF and a standard SMF28 under the same range of input powers, the results are presented in Fig. \ref{fig:6_elm_smf}.
 With a mode field diameter typically twice that of the HNLF, the SMF28 exhibits
 a higher nonlinearity threshold characterized by nonlinear length $L_{\text{NL}}$ after which nonlinear effects appear.
 Importantly, $L_{\text{NL}} = \frac{1}{\gamma P_0}$, where  $\gamma$ is the nonlinear coefficient of the fiber and $P_0 $ is the input peak power.
 In SMF28s, the larger core area reduces $\gamma$, either resulting in a longer $L_{\text{NL}}$, or one needs to compensate by increasing $P_0$.
 As shown in Fig. \ref{fig:6_elm_smf}, the HNLF demonstrates superior performance at moderate input power levels due to its smaller core diameter and higher nonlinear coefficient $\gamma$.
 In contrast, the SMF28 requires significantly higher power to achieve comparable results due to its lower nonlinearity.
 This makes the HNLF a more energy-efficient option for nonlinear optical computing in scenarios where power constraints are critical.
 However, the SMF's robustness at higher power levels, as well as easier optical input coupling, could be advantageous in applications requiring stability over long distances or higher power ranges.

\section{Discussion and Conclusion}

We demonstrated the feasibility of transforming nonlinear optical fibers into a versatile and efficient computational platform.
 Through detailed analysis, it was shown that key fiber parameters, such as dispersion and fiber length, play a pivotal role in defining the computational capacity of the system.
 Using PCA, the dimensionality of the system was evaluated, with results indicating that a range of 70 to 100 PCs can be achieved under optimal conditions for various fibers and input power levels.
 While higher dispersion tends to generate more PCs, potentially enhancing the computational capacity of the platform, it is important to consider the impact of other nonlinear effects.
 In the case of anomalous dispersion, the loss of spectral coherence can undermine repeatability, which we here characterize using the system's consistency.
 Using the fundamental computing measures of dimensionality and consistency, we provide a task-independent characterization and introduce these metrics to the ultrafast nonlinear optics community.
 They enable linking underlying nonlinear dynamical properties to computational capacity, paving the way for systematic analysis of computing via ultrafast NLSE and associated systems in experiments, numerical simulations, and numerically-assisted \cite{Ermolaev24} as well as pure theoretical analysis.

The platform's potential was further validated by successfully classifying the MNIST dataset with 10,000 examples, achieving accuracy levels that significantly exceeded those of a linear classifier and approaches 90$\%$ using only 40 input channels.
 These results underscore the ability of nonlinear optical fibers to exploit their intrinsic physical properties for advanced computational tasks, particularly by leveraging the interplay between nonlinearity, dispersion, and power.

Our optical computing system benefits from the ultrafast nature of light propagation and the ultrafast nonlinear Kerr effect, here leveraging linear as well as nonlinear propagation and mixing at such speeds and bandwidths.
The maximum bandwidth within a single data input example is determined by the system's input pulse width, which in this case is approximately 1.8 THz. This bandwidth can be further increased by using shorter input pulses.

 Combined with the demonstrated computing capability, the real-time classification of ultrafast phenomena on the timescale of the input pulse comes within reach.
 While our current implementation remains constrained by the speed of electronic input devices such as SLMs, we envision a future configuration where the input is provided directly by a real-time data stream, for example, from another femtosecond laser pulse. As the interaction between output weights and their spectral components can be kept passive~\cite{skalli2024annealing}, such a system will be capable of performing non-trivial computations on the original seed pulse, with applications in metrology or novel laser sources. In this context, the system is not positioned as a general-purpose AI accelerator, but rather as a high-seed analog co-processor capable of pulse-to-pulse corrections and metrological tasks.
 This provides prospects for an entirely new class of physical experiments as well as metrology in general.

In summary, this study highlights the significant potential of nonlinear optical fibers as computational platforms, offering a compelling combination of speed and computational capacity.
 Future work will require extensive mapping of the interplay between the different fiber parameters, such as dispersion, input powers as well as the different classes of computational tasks.
 Here, extensive numerical simulations will prove highly beneficial \cite{ermolaev2025limitsnonlineardispersivefiber}.
 Finally, replacing the OSA with a physical output layer and implementing in-situ learning \cite{porte2021complete,skalli2024annealing} will implement a real-time computer with THz input data bandwidth, and inference rate given by the pulse repetition rate and, essentially, negligible latency on the nanosecond scale. A possible implementation uses an SLM and a grating to weight the spectrum optically \cite{skalli2024annealing}. The result can then be read out electronically or kept entirely in the optical domain, preserving fs-scale resolution.

\section{funding}
	The author acknowledge the support of the French Investissements d'Avenir Programme; Agence Nationale de la Recherche (ANR-15-IDEX-0003, ANR-17-EURE-0002, ANR-20-CE30-0004); the Academy of Finland (318082, 320165, 333949); and the European Research Council Consolidator Grant 101044777 (INSPIRE).
	
\section{authorcontributions}
	The author confirms the sole responsibility for the conception of the study, presented result, and manuscript preparation.
	M.H. and P.R. constructed the Fourier-domain pulse shaper used in this work. M.H conducted all experiments.  M.H., D.B., J. M. D. and G.G. planned the research project. All authors participated in interpreting the results and writing the manuscript.

\section{conflictofinterest}
	Authors state no conflict of interest.

\section{dataavailabilitystatement}
	The datasets generated during the current study are available from the corresponding author on reasonable request.

\section*{References}
\bibliography{bibliography}

\end{document}